# Ultrasensitive 1D field-effect phototransistor: $CH_3NH_3PbI_3$ nanowire sensitized individual carbon nanotube

M. Spina[1], ***B. Náfrádi**[1]*, ***H. M. T**óh*áti*[2], K. Kamarás[2], R. Gaal[1], L. Forró[1], E. Horváth[1]*

[1]Laboratory of Physics of Complex Matter (LPMC), Ecole Polytechnique Fédérale de Lausanne, 1015 Lausanne, Switzerland

[2]Institute for Solid State Physics and Optics, Wigner Research Centre for Physics, Hungarian Academy of Sciences, 1525 Budapest, Hungary

**Abstract**: Field-effect phototransistors were fabricated based on individual carbon nanotubes (CNTs) sensitized by $CH_3NH_3PbI_3$ nanowires ($MAPbI_3NW$). These devices represent light responsivities of $R=7.7 \times 10^5$ A/W at low-lighting conditions in the $nWmm^{-2}$ range, unprecedented among CNT-based photo detectors. At high incident power (~1 $mWmm^{-2}$), light soaking results in a negative photocurrent, the device turns insulating. We interpret the phenomenon as a result of efficient electron-hole separation and charge transfer of holes from the perovskite to the carbon nanotube, which improves conductance by increasing the number of carriers, but leaves electrons behind. At high illumination intensity the random electrostatic potential of these quench the mobility in the nanotube. The single CNT device geometry allows the local study of the $MAPbI_3NW$/CNT interface for metallic and semiconducting CNTs separately. Infrared and Raman spectroscopy studies of CNT-$CH_3NH_3PbI_3$ composites revealed that photo-doping takes place at the interface.

In various optoelectronic applications, like light-emitting diodes, photodetectors and photovoltaic cells, semiconducting carbon nanotubes (CNTs) have been successfully used due to their direct band gap and outstanding electronic and mechanical properties.[1] Photodetection of individual CNTs excited by infrared[2-4] (IR) or visible light[5] have been achieved by separating the excitons with large enough electric fields generated locally by asymmetric Schottky contacts[2], p-n junctions[3] or local charge defects[5]. However, their performance has been limited to quantum efficiencies of about 10%[4]. This is mainly limited by the high binding energy and long lifetime of excitons in CNTs.[1,6]

$CH_3NH_3PbI_3$ (MAPbI$_3$) is efficiently used as photosensitizer in many optoelectronic hybrid devices in conjunction with carbon nanomaterials, due to the exceptional, but still not completely explained, physical properties favorable for light harvesting (*i.e.* direct bandgap, large absorption coefficient, long charge diffusion lengths), chemical flexibility and low-cost solution-based processability[7-10]. Several types of heterostructures have been made by combining MAPbI$_3$ and different carbon materials. Fullerenes have been reported to enhance the stability and to reduce drifts and hysteretic effects of MAPbI$_3$ solar cells[11,12]. Incorporation of graphene and carbon nanotube films resulted in semitransparent flexible solar cells[13,14]. The graphene lead-halide interface as a hybrid phototransistor was used as a high-sensitivity phototransistor owing to the successful photo-gating of graphene[15-17].

Here we studied the light induced transfer characteristics of micro-fabricated field-effect transistors, built from individual metallic and semiconducting CNTs and $CH_3NH_3PbI_3$ nanowires (hereafter MAPbI$_3$NW). The sensitization of individual CNT-FETs with a network of MAPbI$_3$ nanowires resulted in responsivities as high as $7.7 \times 10^5$ A/W and external quantum efficiencies of $1.5 \times 10^6$ owing to the successful doping and gating of CNT-FETs. According to our knowledge, our microfabricated hybrid devices attained best-in-class responsivity in low-intensity visible-light detection. The extremely high sensitivity of the present MAPbI$_3$NW/CNT field-effect phototransistors (FEpT) is related to the mechanism reported on graphene/MAPbI$_3$ hybrid photodetectors[15-17]. Importantly, however, because of the unipolar nature of CNT FETs, the present MAPbI$_3$NW/CNT FEpT photodetectors can be switched off unlike the graphene/MAPbI$_3$ counterparts[15-17].

The fabricated field-effect transistors are appropriate tools to obtain valuable information about the light induced charge transfer phenomena at the interface by means of fairly simple electrical transport measurements, as FETs use an electric field to control the conductivity of a channel of one type of charge carrier in a semiconductor material. Detailed analysis of the device characteristics unraveled the charge transfer process between the intimate contact of MAPbI$_3$ and metallic or semiconducting CNTs. Despite the remarkable progress in prototype building, however, there is a lack of knowledge about the fundamental chemical and photo-physical characteristics of the interfaces formed between the carbon nanomaterials and the organometal halide perovskites. Infrared (IR) and Raman spectroscopy of semitransparent highly purified single-walled nanotube buckypapers and MAPbI$_3$ nanowire composites confirmed the observed photo-induced charge transfer process.

**Results and discussion**

The fabrication of our MAPbI$_3$NW/CNT photo-FET has started with the fabrication of an individual CNT based FET. The fabrication of the CNT-FETs involves first the metal catalyst deposition to lithographically predefined positions. The cobalt-containing resist was spin-coated on a highly p-doped Si substrate with 200 nm thick SiO$_2$ thermally grown on top. Patterning the resist by electron-beam lithography created dots of metal ion doped resist as small as 100 nm (Figure 1a). The catalytic nanoparticles were formed by burning the organic resist at 800°C in oxygen (Figure 1b). Carbon nanotubes were grown by CVD at 800°C using ethanol as a carbon source (Figure 1c). Next, CNT-FETs were fabricated by patterning and evaporating the source and drain metal contacts (Ti/Pd 1 nm/70 nm, Figure 1d).

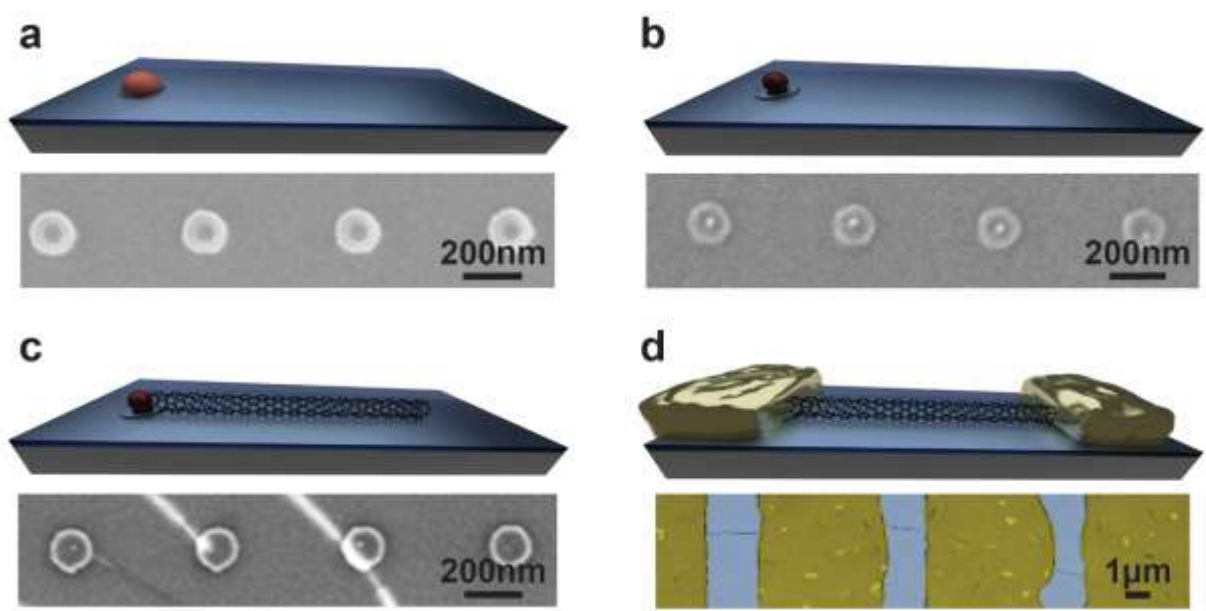

Figure 1. Schematic representation and corresponding false-color SEM micrographs of the process used for synthetizing the CNTs. (a) e-beam lithography patterning of the metal-doped negative-tone resist. (b) Catalyst nanoparticle formation by thermal oxidation. (c) CNT synthesis by ethanol-assisted CVD. (d) Metal contacts deposition by e-beam evaporation.

The CNT-FET (Figure 2a) was sensitized with a network of photoactive MAPbI$_3$ nanowires, deposited by the recently developed slip-coating method[18] (Figure 2b). The hybrid device was subsequently covered with a 500 nm-thick polymethyl methacrylate (PMMA) layer to protect the organometal network from the detrimental effect of humidity.

In the pristine CNT-FET the high work function of the Pd contacts[19] and the p-type doping induced by the exposure to air (O$_2$)[20,21] led to dark transfer characteristics showing unipolar p-type behavior with a threshold voltage $V_{th} \approx 2$ V (Figure 2c). The MAPbI$_3$ nanowire deposition caused both a shift of $V_{th} \approx -1$ V and a decrease of the CNT charge mobility by about 40% (Figure 2c).

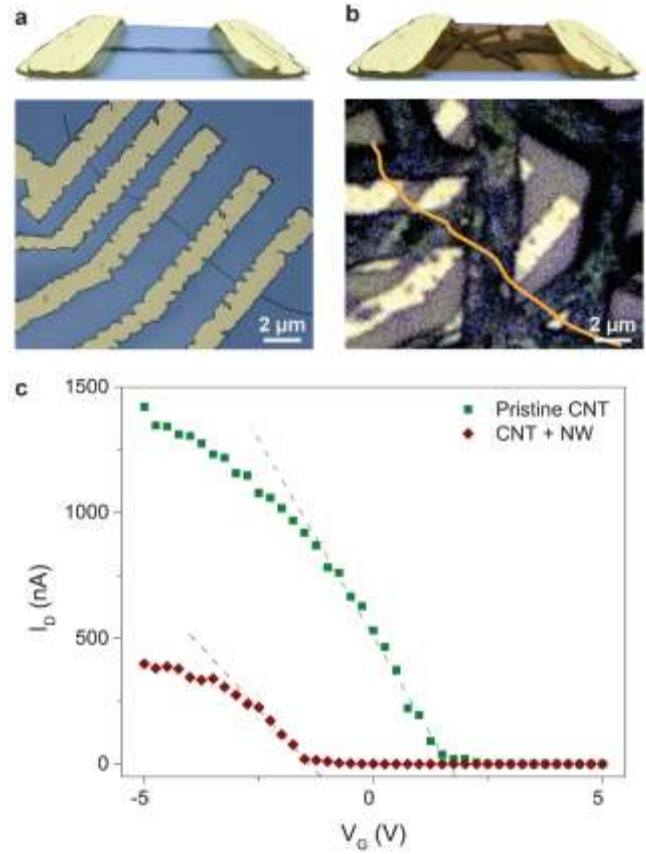

Figure 2: (a) Schematic representation and false color SEM micrograph of a series of a representative CNT-FET. (b) Schematic representation and optical micrograph of a representative MAPbI$_3$NW/CNT-FEpT. (c) Transfer characteristic in dark of a representative device before (green curve) and after (red curve) sensitization ($V_D$=0.2 V). Dashed lines are guide to the eye showing the shift of the on-off threshold voltage $V_{th}$, and the reduction of the CNT mobility upon MAPbI$_3$ deposition.

The central finding of our paper is the remarkable photosensitivity of the hybrid MAPbI$_3$NW/CNT-FEpTs with responsivity $R$=7.7x10$^5$A/W at low light conditions. The photoresponsivity of the hybrid device by the illumination was tested with a red laser ($\lambda$=633 nm) in the 62.5 nWmm$^{-2}$ to 2.5 mWmm$^{-2}$ intensity range. Under illumination, electron-hole pairs are generated in MAPbI$_3$ nanowires. The holes are injected into the nanotube due to the chemical potential mismatch[14], contributing to an increase in the output current in both the ON- and OFF-state of the MAPbI$_3$NW/CNT-FEpTs. Above the threshold voltage, in the OFF-state of the MAPbI$_3$NW/CNT-FEpTs, the photocurrent, hence the total current, $I_D$, of the hybrid device, increases by increasing incident irradiation power (Figure 3b). The $I_D$ current in the OFF-state does not show gate voltage ($V_G$) dependence, thus it corresponds to the intrinsic photocurrent generation of the MAPbI$_3$ nanowire network shortcutting the source-drain contacts, as it was reported in our previous work[18]. The evolution of the ON-state $I_D$ current as a function of illumination intensity and $V_G$ shows a markedly different behavior compared to the OFF-state (Figure 3 and Figure S1,S2). $I_D$-ON shows strong $V_G$ dependence testifying that its origin is predominantly a CNT conduction channel. More interestingly, however, it shows a non-monotonous dependence on illumination intensity (Figure 3 and Figure S1,S2). At low light conditions below 100 nWmm$^{-2}$ intensity $I_D$ increases monotonically by about a factor 2 relative to the dark current. By further increasing the light power, however, $I_D$ rapidly falls and reaches $I_D$ values observed for OFF-state (Figure S1,S2). Illuminating the device with light intensities higher than 95 μWmm$^{-2}$ resulted in a complete switch-off of the nanotube channel conductance over the whole range of positive and negative gate biases applied (Figure 3). At the same time, $V_{th}$ was independent of the light intensity.

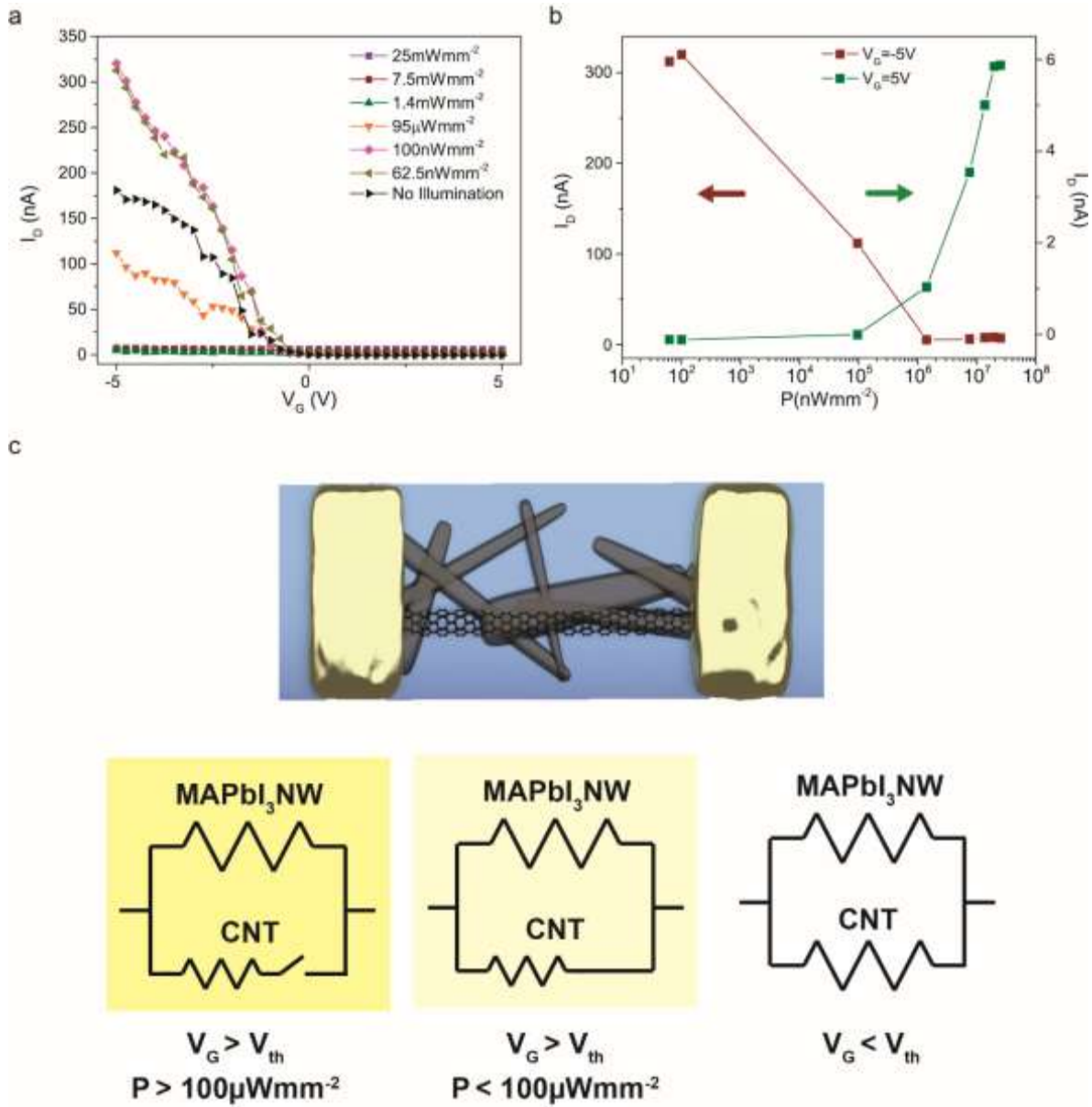

Figure 3. (a) Transfer characteristic of the hybrid phototransistor upon different light irradiation intensities. (b) $I_D$ at $V_G$=5 and -5V as a function of light power. (c) Schematic representation of the two parallel resistor model used to describe our system.

The responsivity ($R$), the magnitude of the electrical signal output in response to a given light power, is one of the most important performance parameters of a photodetector. For the calculation of $R$ the active area of the photodetector is needed. In order to conservatively estimate $R$ of our device, we considered an active area equal to the distance between the contacts (3 μm) multiplied by the carrier diffusion length of photogenerated charge carriers in MAPbI$_3$ reported in the literature (*i.e.* ~1 μm)[22,23]. In the ON-state ($V_G$=-5 V) and at extremely low light intensities (6.25 nWmm$^{-2}$~375 fW) responsivity as high as 7.7x10$^5$ A/W with an external quantum efficiency of $1.5 \times 10^8$ % was measured (Figure 4a). According to our best knowledge this outperforms by about 7 orders of magnitude the best carbon-nanotube based photodetectors reported so far[4]. Moreover, the device gain is highly linear as a function of both $V_D$ (Figure 4a inset) and $V_G$ (Figure 4b) further facilitating applications.

The responsivity in accordance with the device characteristics presented in Figure 3a rapidly drops by increasing the light power and reaches zero when the MAPbI$_3$NW/CNT-FEpTs reaches the light-induced OFF state.

Apart from responsivity, the other important benchmark of photodetector performance is the response time. For the hybrid device the response time to illumination is less than 1 s, (limited by the time resolution of our measurement setup) under all operating conditions tested (Figure 4). On the other hand, the fall-time lasts between ~15 s ($P<95$ μWmm$^{-2}$) and ~35 s ($P>95$ μWmm$^{-2}$).

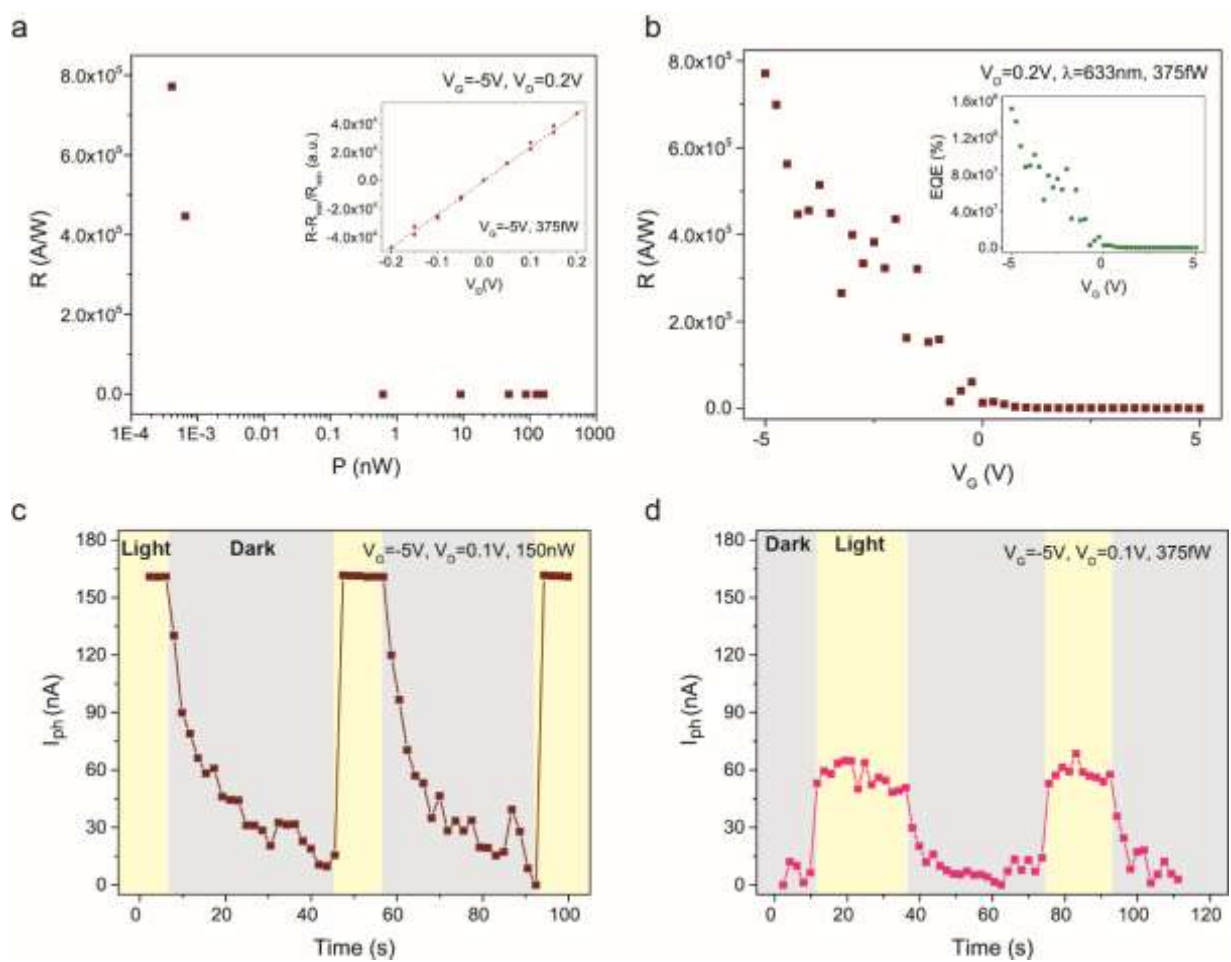

Figure 4. (a) Responsivity of the hybrid device as a function of the light intensity and source-

to-drain voltage (inset). (b) Responsivity and external quantum efficiency under different gate biases. (c-d) Response time under high, 150 nW, (left) and low, 375 fW, (right) light intensity.

The photodiode characteristics of the individual CNT based $MAPbI_3NW/CNT$-FEpT give valuable insight to the interface behavior of CNT and $MAPbI_3NW$. Complications due to the intricate internal behavior of CNT films do not mask intrinsic interface properties, as we use single CNT devices. Moreover, metallic and semiconducting CNTs can be tested separately. In the case of $MAPbI_3NW/CNT$, hybrid devices fabricated from metallic carbon nanotubes the devices showed metallic behavior. Schottky barrier formation was not observed (Figure S2). For semiconducting $CNT/MAPbI_3$ devices the observed shift of $V_{th}$ and the drop of CNT mobility (Figure 2) upon the exposure of CNT-FET to the concentrated $MAPbI_3$-DMF (Dimethylformamide) solution indicate changes of the CNT chemical potential and increased effective disorder along the tubes, respectively.

In order to reveal the potential corrosive effects of the $MAPbI_3$-DMF solution on carbon nanotubes, a free-standing semi-transparent film enriched in semiconducting single walled CNTs was prepared and used as a 3D scaffold for the growth of $MAPbI_3$ nanowires (Figure S3-S5). We studied the interaction of the $MAPbI_3$ nanowires with the CNTs with and without illumination (633 nm LED source) by infrared, near-infrared and Raman spectroscopy. Depositing the nanowires affected only the IR, but not the Raman spectra of the CNT films. (See supporting information Figure S3-S5). We conclude by spectroscopic methods, that under illumination no other significant reaction can be detected between the nanowires and the nanotubes but charge transfer, resulting in mobile carriers. It should be noted, however, that the current running through a functional device can induce ion migration or additional electrochemical redox reactions at the carbon nanotube-$MAPbI_3NWs$ interface which can increase the number of defects, hence reduce the mobility, thus further optical measurements under operating conditions needs to be done to clarify the origin of the mobility drop.

The photodiode characteristics of our $MAPbI_3NW/CNT$-FEpT can be described by a parallel resistor model (Figure 3b). When the hybrid CNT-FET is electrostatically switched off, the resistance of both 1D nanostructures (CNT and $MAPbI_3NW$) are in the GΩ range, the current is low. Due to the closed conduction channel of the CNT, the photocurrent is essentially equal to the photo-generated charges in the $MAPbI_3$ nanowires. If the CNT-FET is in the ON regime due to the electrostatic gating, its high conductivity dominates the performance of the hybrid device. Upon illumination the photo generated positive charges enter the CNT and acts as chemical doping, in agreement with the IR and Raman spectroscopy. The photo-doping, however, does not shift of the chemical potential of the CNTs indicating a vicinity of a Van Hove singularity. The photo-induced negative charges, which are not injected in the CNT due to the work function mismatch[14], constitute scattering centers inhomogeneously distributed along the carbon nanotube. The resulting random potential reduces the charge carrier mobility, thus the overall current. These two effects compete and at high light intensities the detrimental effects of the random potential overcompensate the doping and switch off the CNT conduction channel. Thus the MAPbI$_3$NW/CNT-FEpT acts as a light switch at high powers.

**Conclusion**

In conclusion, we demonstrated gate voltage-dependent visible light photo-response of microfabricated individual MAPbI$_3$NW/CNT photo-FETs for the first time. In the mWmm$^{-2}$ power range light soaking resulted in quenching the conductance of the ON-state p-channel of the individual CNT-FET, and rendering the device to an optical switch. Exposure of these hybrid devices to sub nWmm$^{-2}$ light intensities, however, manifested in strong positive photocurrent. The best devices showed as high as 7.7x10$^5$ A/W responsivity and external quantum efficiencies of 1.5x10$^8$ %, indicating that the device can be used as a low-intensity visible-light detector. We attributed this unconventional photocurrent transfer characteristic of the unique charge distribution over the 1D semiconductor nanotubes. Analysis of the gate dependent transfer characteristics in the dark and under illumination allowed the underlying photon induced charge transfer mechanisms between MAPbI$_3$ and metallic and semiconductor CNTs to be probed. The results have important implications in the fundamental understanding of the photo-physical picture of MAPbI$_3$ and CNT interfaces and in the development and fabrication of organometal halide perovskite based optoelectronic devices as solar cells, LEDs, photodetectors, single photon-detectors and optical switches.

**Experimental Section**

*Resist preparation.* A high-resolution Cobalt-containing negative-tone resist was prepared by dissolving 0.2 wt% of 4-Methyl-1-acetoxycalix[6]arene (Synchem OHG) in monochlorobenzene and 0.2 wt% of Co(III) acetylacetonate, (Sigma-Aldrich GmbH, 99%). After stirring for 1 hour at 700 rpm the solution was filtered through a 0.2-mm Teflon membrane to remove potential solid residues.

*Nanoparticle localization.* The resist was patterned by e-beam lithography with a Vistec EBPG5000 operating at 100kV and 1nA. The nucleation centers were localized by a reactive ion etch step of 10 seconds with an Adixen AMS200 and a gas mixture of Ar and $C_4F_8$.

*Carbon Nanotube synthesis.* The deposited nanoparticles are catalytically activated by a 10 min reduction at 800 °C under controlled atmosphere (Ar/$H_2$ 8:1 vol%). Next, ethanol vapor was introduced in the quartz tube using Argon and Hydrogen (1:2 vol%) as carrier gas. After 5 minutes the carbon source was evacuated and the samples were cooled down to room temperature.

*Carbon Nanotube film synthesis.* Films of single walled carbon nanotubes were prepared from P2 and semiconductor enriched nanotubes as described by Borondics et al.[24]

*$MAPbI_3$ nanowires synthesis.* The network of $MAPbI_3$ nanowires was subsequently deposited by the slip-coating technique reported by Horv th et al.[18]

*Photoelectrical characterization.* The photoelectric response measurements of the fabricated hybrid devices were performed using a standard DC technique. The light sources used were a red laser beam ($\lambda$=633 nm) with a spot size of about 4 mm. All the measurements were performed at room temperature and in ambient environment.

**Conflict of Interests:** The authors declare no competing financial interest.

**Acknowledgment.** The support of the Swiss national science foundation is gratefully acknowledged. The work in Budapest was supported by the Hungarian National Research Fund (OTKA) No. 105691.

**Supporting Information Available**: Infrared and Raman spectroscopy with additional electronic transfer characterizations are included. This material is available free of charge *via* the internet at http://pubs.acs.org

# Supporting Information

# Ultrasensitive 1D field-effect phototransistor: $CH_3NH_3PbI_3$ nanowire sensitized individual carbon nanotube

M. Spina[1], *B. Náfrádi[1]*, *H. M. Tóháti[2]*, K. Kamarás[2], R.Gaal[1], L. Forró[1], E. Horváth[1]*

*Estimation of the charge carrier mobility*

The hole mobilities of CNT-FETs were extracted from the linear region of the transfer characteristics using the expression for the low-field field-effect mobility

(Equation 1)

where $C_i$ is the capacitance of the gate insulator (= $\varepsilon_0\varepsilon_r/d$ = 1.64 $10^4$ F m$^{-2}$).

*Transfer characteristic in the light induced off state*

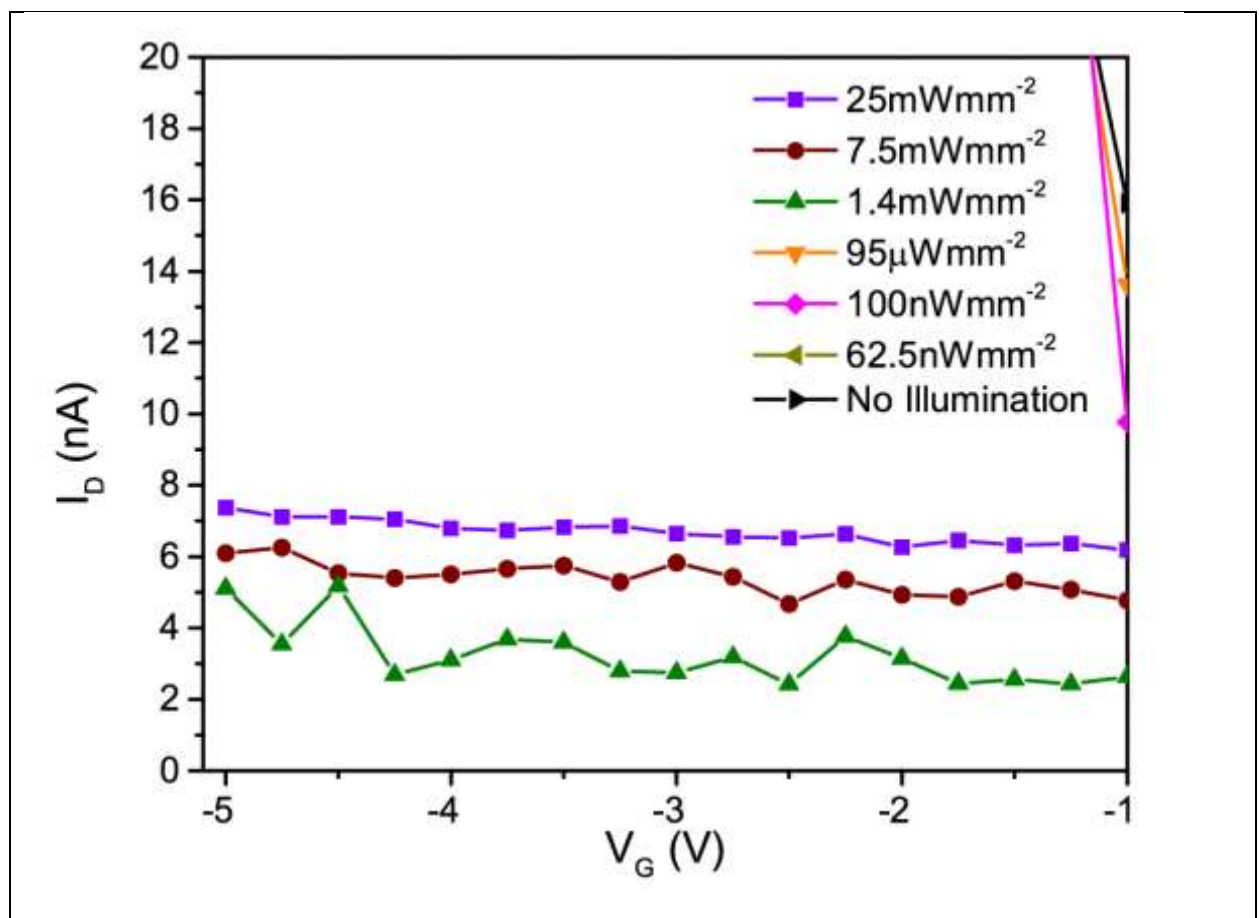

*Figure S1. Transfer characteristic of the hybrid phototransistor presented in the main text upon different high light irradiation intensities in the electrostatic ON-state.*

*Transfer Characteristics of additional Hybrid Devices*

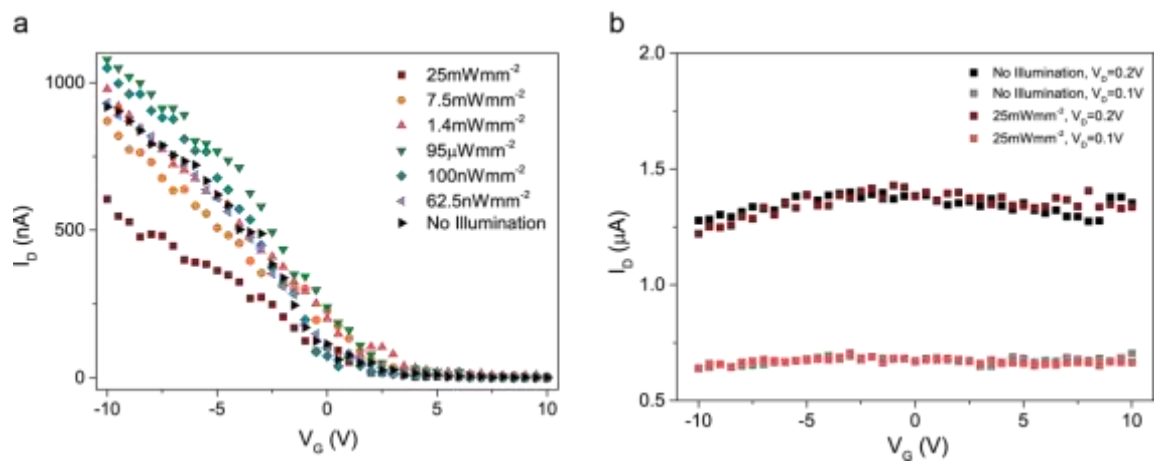

*Figure S2. Transfer characteristic of a semiconducting (a) and a metallic (b) CNT-FET/MAPbI$_3$NW hybrid device.*

An additional semiconducting hybrid device was characterized under the same experimental conditions as the one described in the main text. The photo-induced charge modulation (both positive and negative) was also recorded, with the difference that the number of scattering centers generated was not enough to completely switch off the channel (Figure S1a).

Moreover, an individual metallic nanotube was also sensitized with the perovskite nanowires but in this case no photocurrent was observed (Figure S1b).

*Infrared and Raman spectroscopy*

Depositing the $CH_3NH_3PbI_3$ nanowires affected only the IR, but not the Raman spectra of the CNT Buckypaper films. As the strong resonance Raman lines of the nanotubes dominate the spectra over molecular vibrations (Figure S2), information on chemical transformation is drawn from the D/G peak ratio. Chemical functionalization should result in the increase of this ratio; the lack of such increase indicates that little, if any, change in the bonds attached to the tube walls occurred. Charge transfer, *i.e.* doping of the tubes would cause an overall decrease in all peak intensities; however, such a change is difficult to observe because of the relative nature of the measurement (all spectra are normalized to the G peak) and the necessity of baseline correction due to the strong $MAPbI_3$ luminescence. Such subtle effects in the electronic structure show up much more clearly in the IR/NIR spectra.

After initial deposition of the nanowires, infrared-active vibrations of $MAPbI_3$[1] appeared (Figure S3) but the $S_{11}$ electronic transition of the carbon nanotubes was not affected (Figure S4). Significant changes happened on illumination, as shown in Figure S4. The quantity derived from the measured transmittance, $T$, is

$$\overline{\qquad\qquad},\qquad\qquad\text{(Equation 2)}$$

the change in absorption due to illumination. This change consists of an increase in the low-frequency absorption, together with a decrease in the $S_{11}$ transition, indicative of charge transfer[2]. Most of the increase is observed in the first 10 minutes of illumination, but it is continuous up to about 35 minutes. The effect is reversible, although relaxation to the original state after switching off the light is slower.

*Raman spectra*

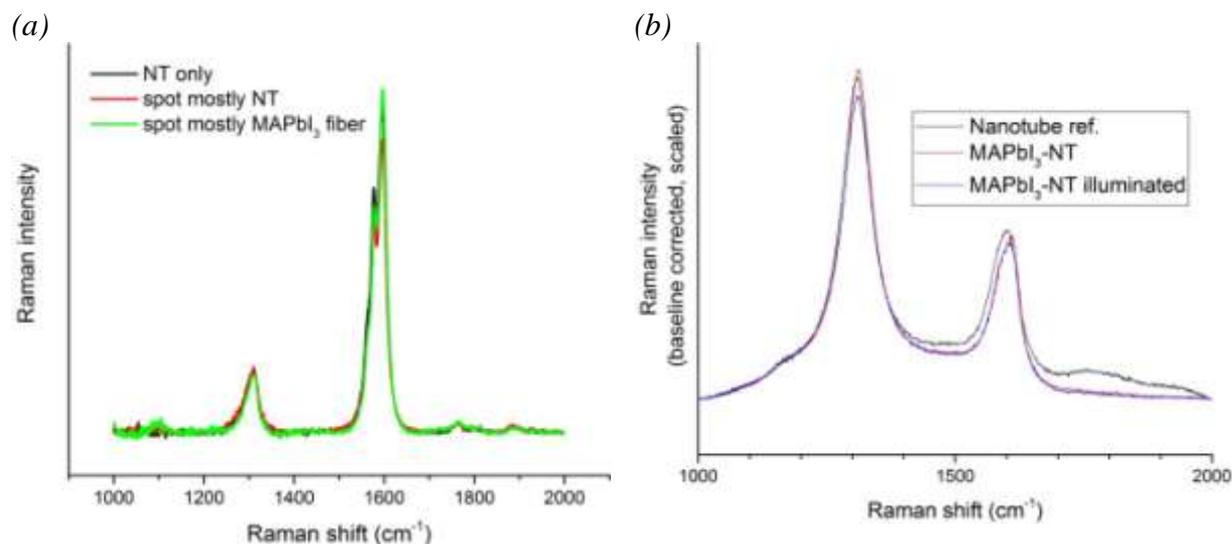

*Figure S3 (a) Raman spectra of mixed single-walled nanotubes before and after treatment with MAPbI$_3$. Spectra taken close to the MAPbI$_3$ fibers and at untreated parts of the sample did not show significant difference in D/G ratio. Spectra are normalized to the G band.*
*(b) Raman spectra of mixed multi-walled nanotubes before and after treatment with MAPbI$_3$, and after illumination with 633 nm light. Neither treatment with MAPbI$_3$ fibers nor illumination resulted in any significant change in D/G ratio. Spectra are normalized to the G band.*

*Infrared Spectra*

Below we compare the infrared spectra of MAPbI$_3$ pristine nanowires with those of the hybrid structures formed by MAPbI$_3$NW and CNT films. Adhesion to the surface splits some lines in the MAPbI$_3$ spectrum. The nanowire spectra were recorded in diffuse reflectance (DRIFT) mode, all others were calculated from transmission.

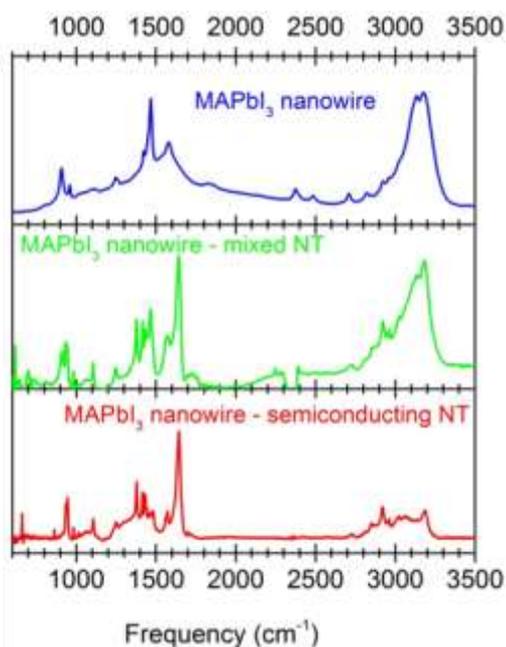

*Figure S4 Infrared spectra of MAPbI$_3$ as-prepared nanowire, and hybrids from both mixed and semiconducting single-walled nanotube films. Absorption was calculated from diffuse reflectance (DRIFT) for nanowires and from transmission in all other cases. Baseline correction was applied to absorbance spectra.*

*Changes in infrared spectra upon illumination*

Based on IR spectra below we compare the change of a MAPbI$_3$-semiconducting CNT hybrid to that of the pristine nanotube sample after 10 minutes of illumination. The hybrid structure shows the effects of nanotube doping (explained in the DOS illustrations for electron doping): increase in the low-frequency free-carrier absorption and decrease of the S$_{11}$ transition intensity due to filling of final states.

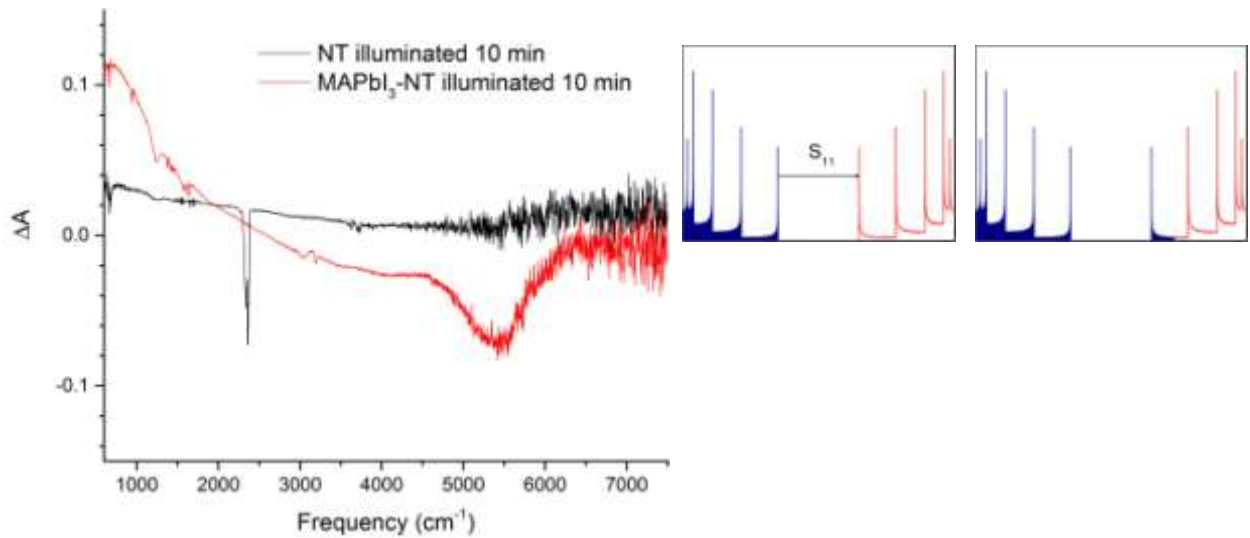

*Figure S5 Change in absorption upon 10 minutes illumination by 633 nm light of a semiconducting nanotube film (black curve) and a MAPbI$_3$-semiconducting SWNT hybrid (red curve). The red curve clearly shows the effects of added carriers into the CNT (illustrated on the right). The sharp peaks in the nanotube spectrum are due to atmospheric carbon dioxide.*